# Carrier-induced Phase Transition in Metal Dichlorides XCl$_2$ (X: Fe, Co, and Ni)

**Teguh Budi Prayitno**

Department of Physics, Faculty of Mathematics and Natural Science, Universitas Negeri Jakarta, Kampus A Jl. Rawamangun Muka, Jakarta Timur 13220, Indonesia

Teguh-budi@unj.ac.id

**Abstract**

We investigated the ground state of monolayer 1T-XCl$_2$ (X: Fe, Co, and Ni) using the generalized Bloch theorem, which can generate ferromagnetic, spiral, and antiferromagnetic states. Each state was represented by a unique spiral vector that arranges the magnetic moment of magnetic atom in the primitive unit cell. We found the ferromagnetic ground state for the FeCl$_2$ and NiCl$_2$ while the spiral ground state appears for the CoCl$_2$. We also showed that the ground state depends sensitively on the lattice constant. When the hole−electron doping was taken into account, we found the phase transition, which involves the ferromagnetic, spiral, and antiferromagnetic states, for all the systems. Since the spin-spin interaction in the monolayer metal dichlorides is influenced by the competition between the direct exchange and the superexchange, we justify that the carrier concentration determines which interaction should dominate.

## 1. Introduction

Recently, transition metal dihalides XY$_2$ (X: 3$d$ transition metal cation and Y: halogen anion) give a great attention on magnetism due to electric and magnetic properties, such as multiferroic property [1-3], which is potential for the future spintronic devices [4]. In addition, various ground states emerge for any structures that make the metal dihalides show the potentiality for designing and developing metal-dihalide-based spintronics [5, 6]. However, the experiment verification is only available for the bulk structure [7, 8]. For the low-dimensional version, the ground state is only predicted by the density functional theory (DFT).

Explorations on magnetism in two-dimensional (2D) materials become more intensively considered since the discovery of graphene by Novoselov and his co-workers [9-11] for future applicable electronic devices. In 2D metal dihalides, there are several reports that these materials show the properties as a semiconductor [12], topological insulator [13], and half metal [14]. Nevertheless, the ground states for some 2D metal dihalides meet discrepancy. Previous works only considered two different ground states, namely, the ferromagnetic (FM) and antiferromagnetic (AFM) states [15, 16], but never mentioned the helimagnetic (HM) state or spiral (SP) state. At the same time, the prediction of HM state or SP state was reported in Refs. [6, 17]. Note that the SP state is a typical case of HM state where the cone angle to form the SP state should be fixed in the calculation [18].

In some materials especially in the metal dihalides [3], the HM state or SP state can be utilized to model a domain wall [19-22] for analyzing the physical properties such as magnetoresistance and ferroelectricity. The interesting case of the appearance of SP state in 2D metal dihalides, in this case, that it does exist without any physical or chemical treatments, thus the SP state appears in the stable form. For the comparison, zigzag graphene nanoribbon as a lower-dimensional form of graphene shows an SP state when the electric field is employed [23]. To construct an SP formation, indeed, it requires a very large cell, thus needing a high computational cost. In addition, the calculation sometimes is very difficult to converge due to the non-collinear system with various different magnetic moments.

The purpose of this paper is to investigate the ground state of monolayer metal dichlorides 1T-$XCl_2$ (X: Fe, Co, and Ni), kinds of metal dihalides, using the generalized Bloch theorem (GBT) approach without spin-orbit interaction in the primitive unit cell, which contains one magnetic atom X as a cation and two non-magnetic atoms Cl as the anions. In contrast to the supercell approach, the GBT can predict the SP state even for the small spiral vector which needs a very large cell to produce the same configuration, thus the calculation can be performed efficiently. In this case, we only consider three stable states, namely, the FM, SP, and AFM states within the self-consistent non-collinear DFT. Since the SP state is a manifestation of the frustrated spin due to the competition between the direct exchange and the superexchange, its spin-spin interaction is determined by the energy difference between the FM and AFM states.

We observe the FM ground state for $FeCl_2$ and $NiCl_2$, and an SP ground state for $CoCl_2$. All the related references reported the same ground state for $FeCl_2$ and $NiCl_2$, but never predicted the SP state for $CoCl_2$. Besides, we also find the other discrepancy in the $CoCl_2$, i.e., the AFM state is more stable than the FM state whereas the FM state should be more stable [15]. We justify that the discrepancies may be caused by the lattice constant. When the lattice constant is smaller than the experimental lattice from bulk structure, we find the dissimilar results with our previous calculations, but in a good agreement for the ground state of $FeCl_2$ and $CoCl_2$ as in Ref. [15]. In addition, as for $NiCl_2$, we find an SP ground state instead of the FM state, contrary to our previous calculation.

We also consider the hole−electron doping to see the existence of the phase transition. Compared to Ref. [24], even though only one $X^{2+}$ cation in the unit cell, taking the hole−electron doping into account can be realized in the experiment, for example by defect [25]. When we enhance the concentration, we observe the phase transition, thus indicating the competition of spin-spin interaction due to carrier concentration. This means that the doping can induce the other ground states, which can be important for the applicable spintronic devices.

## 2. Computational Details

The detailed self-consistent calculations were carried out by the OpenMX code [26] with the norm-conserving pseudo-potentials approach [27]. In the code, the Bloch wavefunction, which employs the GBT, is represented by the linear combination of pseudo-atomic orbital (LCPAO) as [28]

$$\psi_{\mu\mathbf{k}}(\mathbf{r}) = \frac{1}{\sqrt{N}} \sum_n^N \sum_{j\alpha} \left[ e^{i(\mathbf{k}-\mathbf{q}/2)\cdot\mathbf{R}_n} C_{\mu\mathbf{k},j\alpha}^{\uparrow} \begin{pmatrix} 1 \\ 0 \end{pmatrix} + e^{i(\mathbf{k}+\mathbf{q}/2)\cdot\mathbf{R}_n} C_{\mu\mathbf{k},j\alpha}^{\downarrow} \begin{pmatrix} 0 \\ 1 \end{pmatrix} \right]$$
$$\times \phi_{j\alpha}(\mathbf{r} - \tau_j - \mathbf{R}_n) \tag{1}$$

where $\mathbf{q}$ and $\phi_{j\alpha}$ denote the spiral vector and localized orbital function. This orbital function is numerically produced by the confinement scheme with the boundary condition [29, 30]. To ensure the converged results for all $\mathbf{q}$ in the non-collinear calculations, we set the generalized gradient approximation (GGA) for the exchange interaction [31] with a $20 \times 20 \times 1$ k-point sampling in the Brillouin zone and a cutoff energy of 200 Ryd. The non-magnetic condition was applied to set the atomic positions where the acting force on atom is less than 0.0001 Hartree/Å.

We used a primitive two-dimensional monolayer 1T structure of metal dichlorides $XCl_2$ as shown by a parallelogram in Fig. 1, in which the metal cations X (Fe, Co, and Ni) and anions Cl are drawn by the large and small spheres, respectively. For the basis set of metal cation atoms, we assigned three *s*, three *p*, three *d*, and two *f* primitive valence orbitals within the cutoff radius of 4.0 Bohr. Meanwhile, two *s*, two *p*, and one *d* primitive valence orbitals within the cutoff radius of 7.0 Bohr were assigned for Cl atoms. All the lattice constants were taken from the experiments of the bulk structure, namely 3.57 Å for FeCl$_2$, 3.54 Å for CoCl$_2$, and 3.48 Å for NiCl$_2$ [15] with the same space group $R\bar{3}m$. We set the vacuum distance of 17.47 Å to generate a vacuum in the non-periodic cell. However, we will see in the next section, that increasing the vacuum distance leads to the different energy scale if we do not increase the number of orbitals.

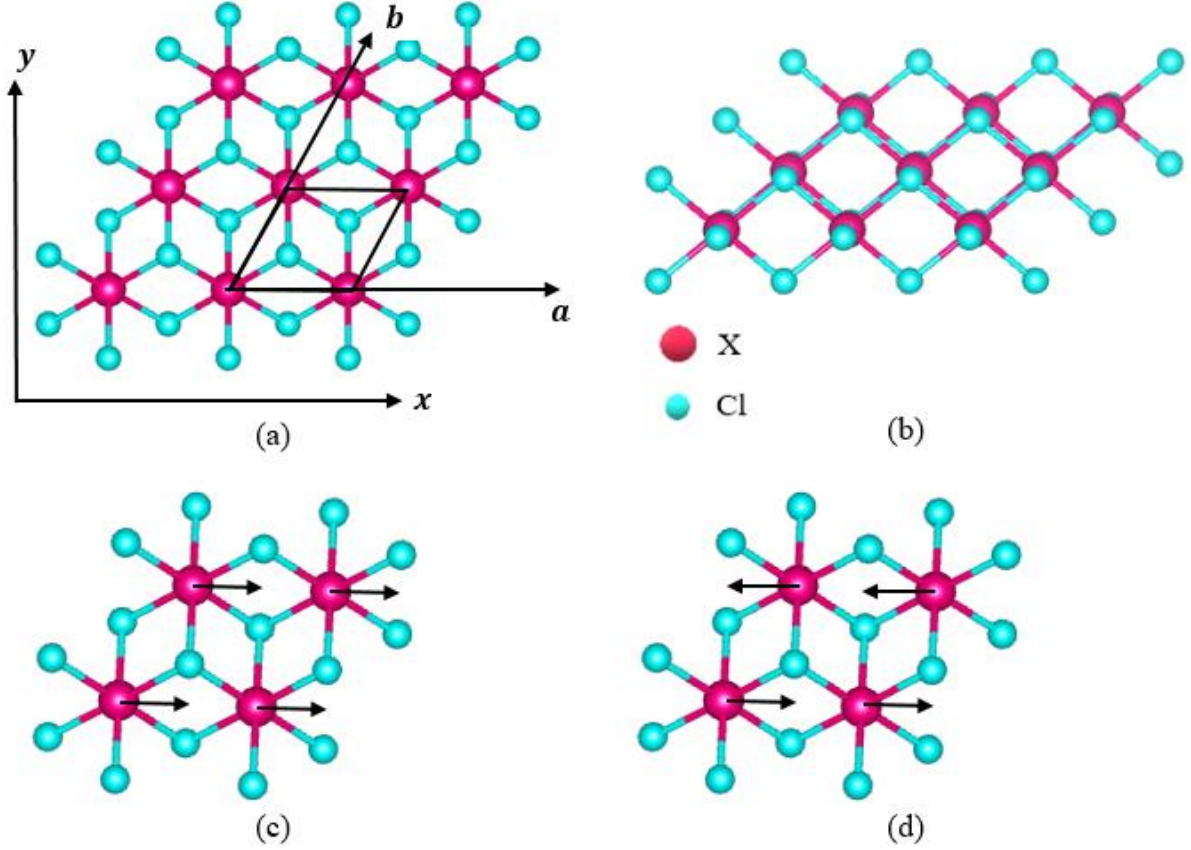

**Figure 1**. (Color online) Top view (a) and side view (b) of crystal structures of 1T monolayer dichlorides XCl$_2$ (X: Fe, Co, and Ni). FM structure (c) and AFM structure (d) in primitive unit cell are generated by $\phi = 0$ and $\phi = 1$, respectively. The Primitive unit cell is shown by the parallelogram in (a).

In this paper, we only consider three states, i.e., the FM, AFM, and SP states. To realize it, we develop a planar spiral configuration ($\theta = 90°$) where the magnetic moment **M** of metal atoms X will be rotated with the condition [32]

$$\mathbf{M}_i = M_i \begin{pmatrix} \cos(\mathbf{q} \cdot \mathbf{R}_i) \sin\theta_i \\ \sin(\mathbf{q} \cdot \mathbf{R}_i) \sin\theta_i \\ \cos\theta_i \end{pmatrix}. \tag{2}$$

As shown in Fig. 1(a), in terms of the lattice constant $a$, we specify the lattice vectors [17]

$$\boldsymbol{a} = a\hat{e}_x, \qquad \boldsymbol{b} = \frac{a}{2}\hat{e}_x + \frac{a}{2}\sqrt{3}\hat{e}_y, \tag{3}$$

and the reciprocal lattice vectors

$$\boldsymbol{A} = \frac{2\pi}{a}\hat{e}_x - \frac{2\pi}{a\sqrt{3}}\hat{e}_y, \qquad \boldsymbol{B} = \frac{4\pi}{a\sqrt{3}}\hat{e}_y. \tag{4}$$

To accommodate the three states, we choose the direction of **q** as

$$\mathbf{q} = \phi(\mathbf{A} + 0.5\mathbf{B}) = \phi\frac{2\pi}{a}\hat{e}_x. \tag{5}$$

In Eq. (5), $\phi$ is a set of real numbers between 0 and 1 where the FM and AFM states are generated by setting $\phi = 0$ and $\phi = 1$ respectively, as shown in Figs. 1(c) and 1(d). At the same time, the SP state is generated by setting $\phi$ between 0 and 1.

## 3. Results and Discussions
### 3.1 Non-doped case

As shown in Fig. 2(a), the ground states of $FeCl_2$ and $NiCl_2$ are the FM state since their lowest energies lie in $\phi = 0$. Meanwhile, the $CoCl_2$ has an SP state where the lowest state is in $\phi = 0.48$, see Fig. 3 for the illustration. Here, the directions of magnetic moments of Co atoms obey Eq. 2. Based on our calculations, the FM ground state for the $FeCl_2$ and $NiCl_2$, are the same results with the prediction in Refs. [15, 16], but not for the $CoCl_2$. As for the $CoCl_2$, they never predict the HM state or SP state, thus our result gives the appearance of SP state in the $CoCl_2$. In addition, we also see the nearly same trends for the magnetic moments for all systems as shown in Fig. 2(b), namely, all the magnetic moments for all $\phi$ are almost the same, where the difference between the highest and lowest magnetic moment for all the systems is less than 0.15 $\mu_B$. However, this trend is not the same for the monolayer zigzag graphene nanoribbon where the magnetic moment reduces rapidly as $\mathbf{q}$ increases [33]. Besides, compared to the fcc iron case where the difference between those two magnetic moments is almost 1 $\mu_B$ [34-37], the magnetic moments for monolayer 1T-$XCl_2$ are more robust to $\mathbf{q}$.

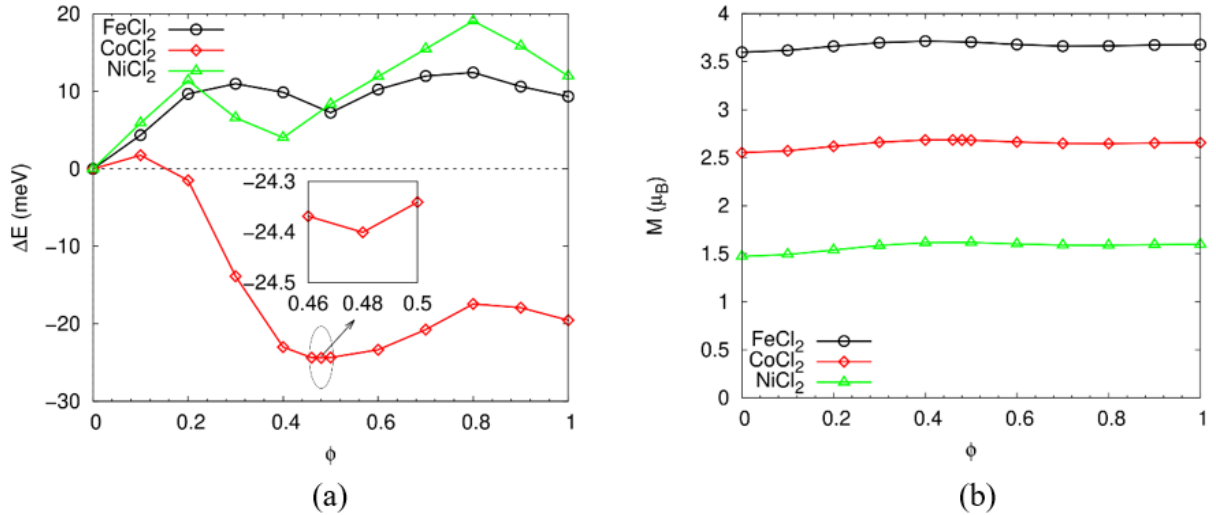

**Figure 2**. (Color online) Total energy difference (a) and magnetic moment per magnetic atom (b) as the function of $\phi$ along *x*- direction for the non-doped case with the experimental lattice constants and vacuum distance of 17.47 Å.

When the exchange parameter $J_{ij}$ was used to determine the kind of interaction, Ref. [15] defined the exchange parameter as an energy difference between the AFM and FM states. This parameter in the monolayer dichlorides is determined by the spin-spin competition of the AFM direct exchange against the FM superexchange as illustrated in Fig. 4, where both of them rely on the electron hopping from an X site to its nearest neighbour X site. Indeed, this hopping should obey the Hund's rule through the Coulomb repulsion. If the $J_{ij}$ is positive, the magnetic order favors the FM state, thus the FM superexchange wins against the AFM direct exchange, otherwise, the magnetic order prefers the AFM states when the AFM direct exchange is more dominant than the FM superexchange. In this condition, the SP state may be caused the spin frustration when the spin-spin interaction must choose either the FM superexchange or the AFM direct exchange.

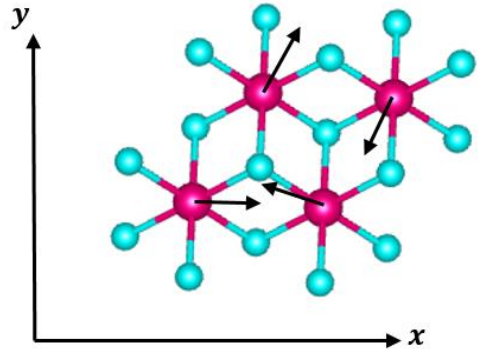

**Figure 3**. (Color online) Spiral magnetic configuration for CoCl2 at $\phi = 0.48$.

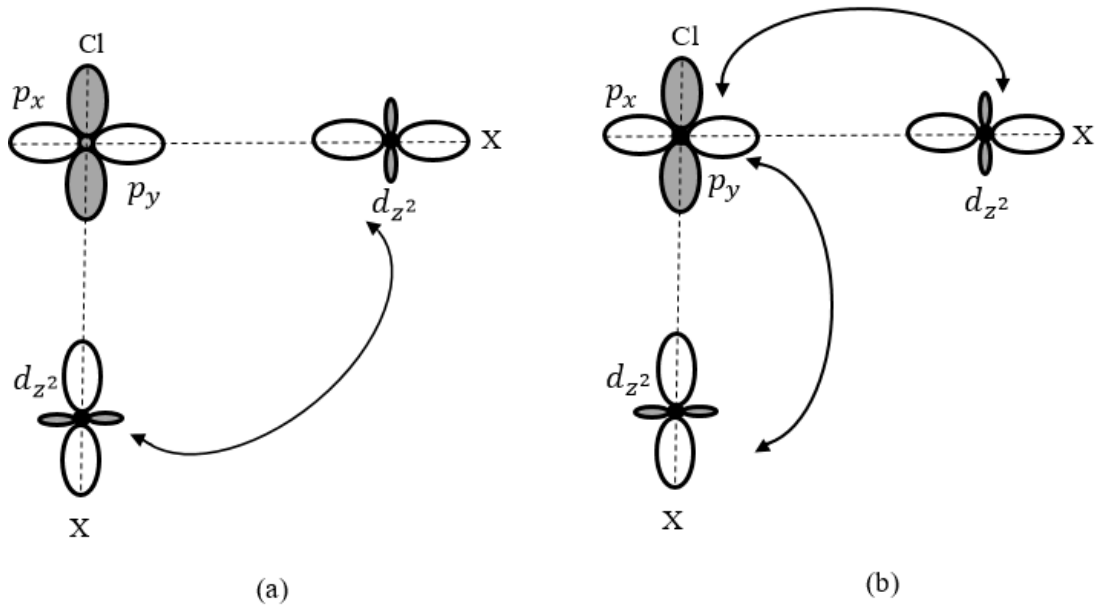

**Figure 4**. Illustration of AFM direct exchange (a) and FM superexchange (b).

The FM superexchange in this system is well deduced from the Goodenough–Kanamori–Anderson (GKA) rules [38-40]. According to the rules, the FM superexchange is preferred if the angle of metal–anion–metal is $90°$. Meanwhile, if the angle of metal–anion–metal is $180°$, the AFM superexchange takes place. Since the angle of X–Cl–X is close to $90°$, according to the GKA rules, the interaction in our systems should be an FM superexchange. Note that the FM superexchange is usually weaker than the AFM superexchange due to the bonding between $p$ and $d$ orbitals. As for the FeCl$_2$ and NiCl$_2$, the FM state is more stable than the AFM state, in good agreements with Refs. [6, 15]. Therefore, the FM superexchange overcomes the AFM direct exchange. However, we obtain different result for the CoCl$_2$, namely, the AFM state is more stable than the FM state, thus the AFM direct exchange dominates the magnetic order. When we investigate the cause of why this discrepancy in the CoCl$_2$ occurs, we find that the magnetic order is determined sensitively by the lattice constant.

The lattice constants used in Ref. [15] are 3.43 Å for $FeCl_2$, and 3.42 Å for both $CoCl_2$ and $NiCl_2$, which were obtained by the optimization. Our calculations using the above lattice constants are not still in good agreements with those when we set the vacuum distance close to 17 Å, i.e. 20 Å as shown in Fig. 5(a). At last, when we enhance the vacuum distance up to 30 Å as shown in Fig. 5(c), we obtain that the FM state is more stable than the AFM state for all systems as reported in Refs. [6, 15]. Here, we observe the change of the ground state in $CoCl_2$ and $NiCl_2$. While the FM ground state still holds for the $FeCl_2$, the FM and SP states refer to the $CoCl_2$ and $NiCl_2$, respectively. Therefore, we strongly justify that the ground state in the monolayer dichlorides can also depend on the lattice constant, which will be proven in the next discussion. Note that the magnetic moments seem also robust to the vacuum distance, as shown in Figs. 5(b) and 5(d). Based on our results, if we consider the experimental lattice as the reference, the smaller lattice constants obtained by the optimization meet the agreements as long as the vacuum distance is quite large. However, when we still use the experimental lattice constant, we never meet the agreement even for increasing the vacuum distance up to 30 Å.

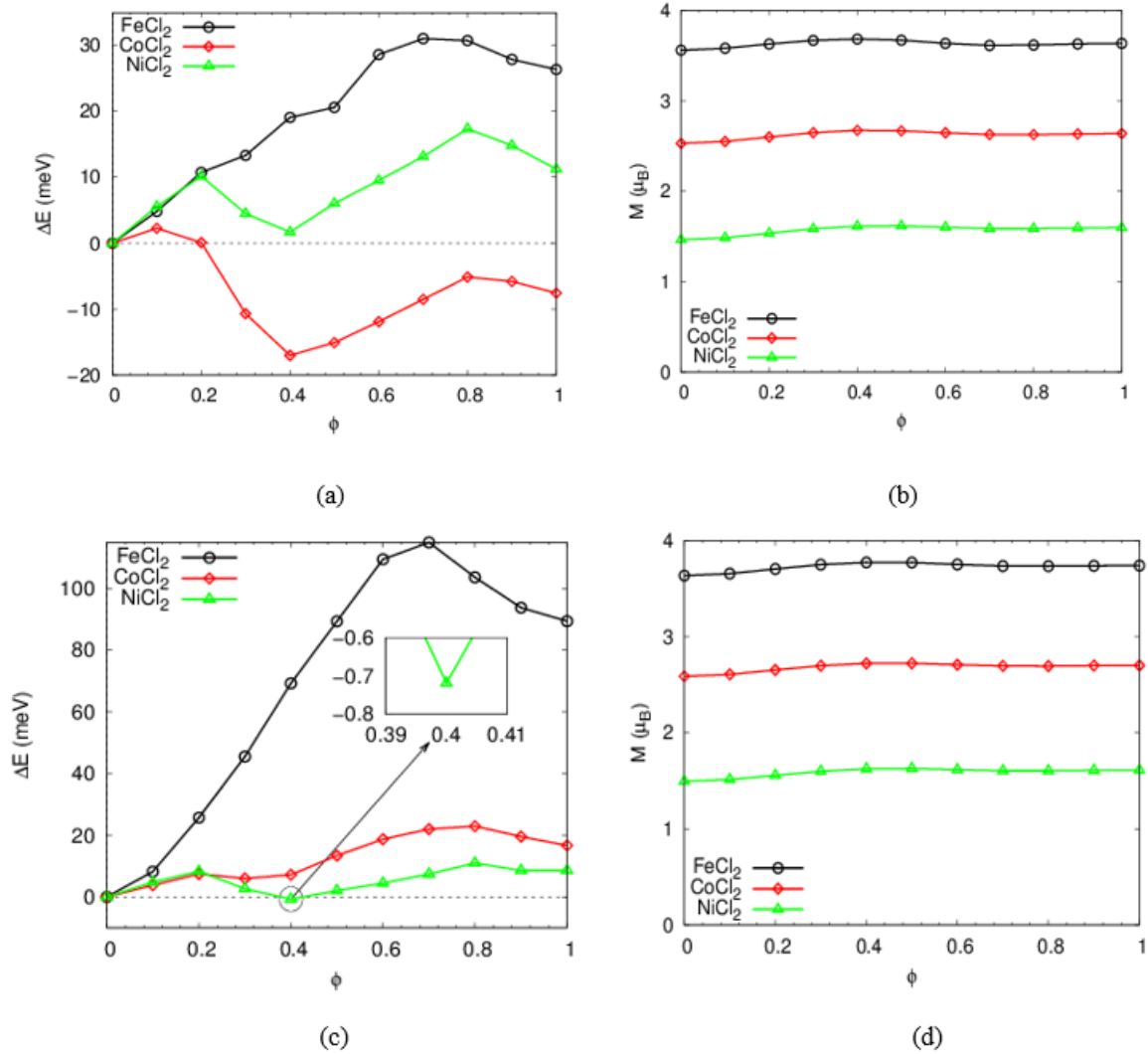

**Figure 5**. (Color online) Total energy difference and magnetic moment per magnetic atom as the function of $\phi$ along $x$- direction for the non-doped case with vacuum distance of 20 Å ((a) and (b)) and 30 Å ((c) and (d)). Here, the lattice constants are taken from Ref. [15].

When we plot the dependence of energy difference $\Delta E$ between the AFM ($\phi = 1$) and FM ($\phi = 0$) states on the vacuum distance, we observe that all the $\Delta E$ are so sensitive to the vacuum distance especially for the large vacuum distance. As shown in Fig. 6(a), all the $\Delta E$ are not slightly different for the small distance while the significant differences appear in the large vacuum distance. However, the dependence of $\Delta E$ on the vacuum distance can be reduced by increasing the number of orbitals, as shown in Fig. 6(b). Our previous calculations use 54 orbitals. When we increase the number orbitals up to 58 orbitals, the dependence becomes insignificant while the energy difference significantly changes. We also observe that the FM state is more stable than the AFM state for all the systems as reported earlier in Ref. [15]. This means that the additional number of orbitals is required to remove the dependence of the vacuum distance and to obtain the more reliable results. However, one should be careful when increasing the number of orbitals as large as possible due to an overcompleteness problem. In Ref, [29], an overcompleteness problem may cause unreliable energy due to negative eigenvalue in the self-consistent calculations.

Comparing Figs. 2 and 5, we also suspect that the magnetic ground states should depend on the lattice constant, too. To investigate it, we plot the total energy differences for four lattice constants for all the systems, as shown in Fig. 7. Except the $FeCl_2$ in Fig. 7(a), we observe various ground states in the $CoCl_2$ and $NiCl_2$ as the lattice constant increases, as shown in Figs. 7(c) and 7(e) respectively. For the small lattice constants, all the systems tend to either the FM state or the SP state, while the AFM state appears in the large lattice constant except the $FeCl_2$. The main reason why the FM state preserves in $FeCl_2$ is due to the half-metallic property while the $CoCl_2$ and $NiCl_2$ show the insulating property [16]. Regarding to the lattice-dependent ground state, we predict that the phase transition may also be generated by strain. We also show that all the magnetic moments do not depend sensitively on the lattice constant, as shown in Figs. 7(b), 7(d), and 7(f). Based on the spin-spin competition, we conclude that the FM superexchange works well in the small lattice constant while the AFM direct exchange takes place in the large lattice constant.

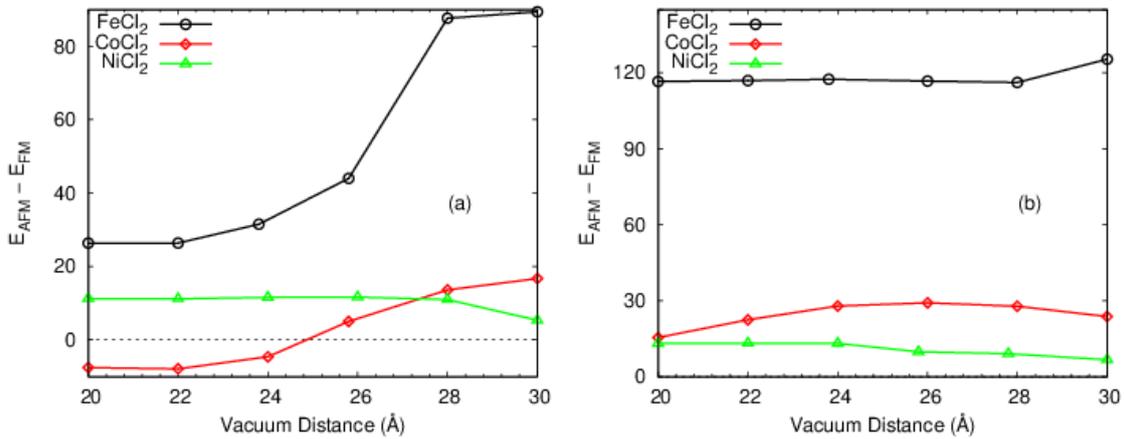

**Figure 6**. (Color online) Dependence of energy difference $\Delta E$ between AFM state ($\phi = 1$) and FM state ($\phi = 0$) on the vacuum distance for 54 orbitals (a) and for 58 orbitals (b). Here, the lattice constants are taken from Ref. [15].

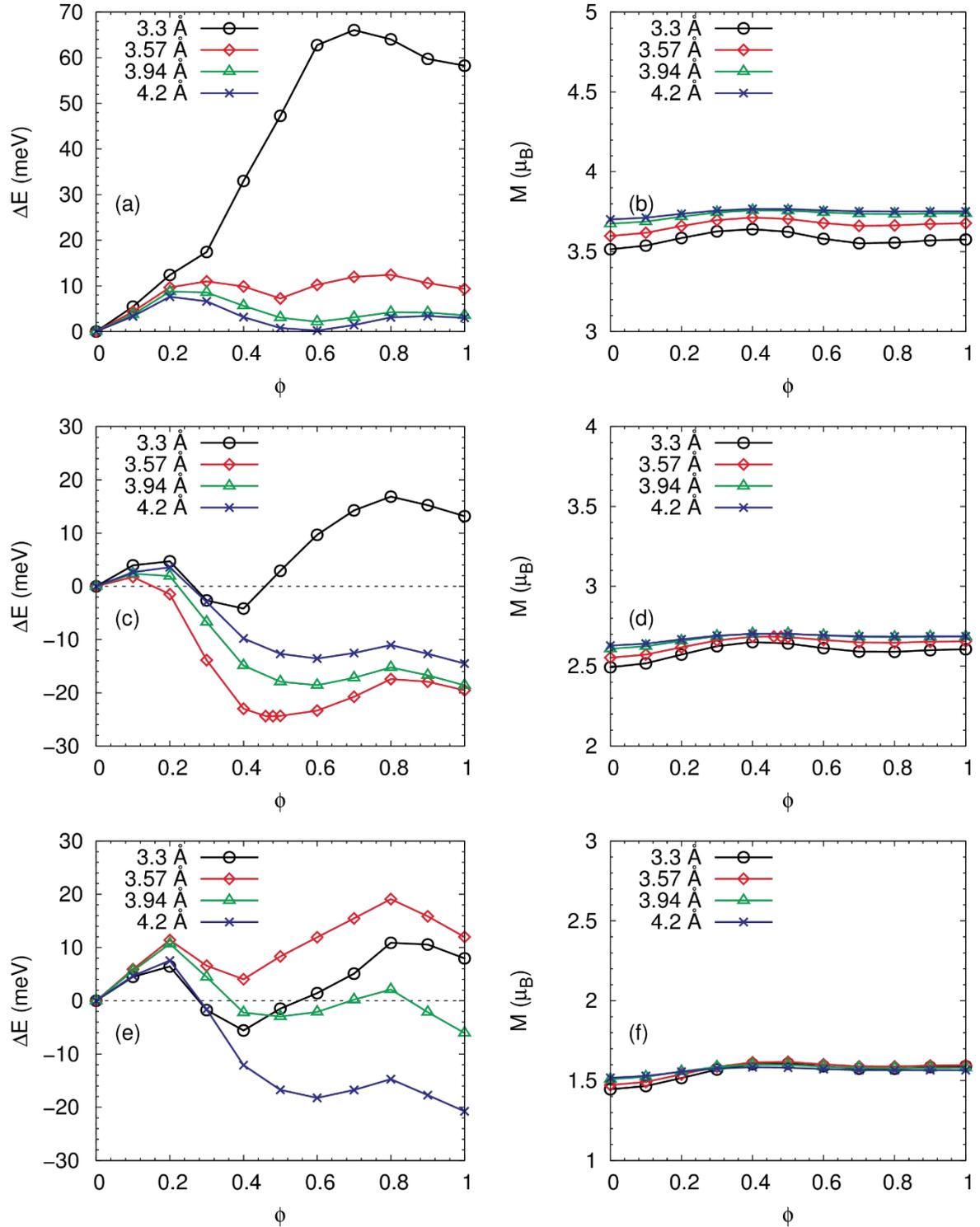

**Figure 7**. (Color online) Various ground states and related magnetic moments in terms of lattice constant for FeCl$_2$ ((a) and (b)), CoCl$_2$ ((c) and (d)), and NiCl$_2$ ((e) and (f)), with the vacuum distance of 17.47 Å.

**3.2 Doped case**

All the self-consistent calculations in the carrier doped monolayer 1T-XCl$_2$ is carried out by the Fermi-level shift (FLS) method. In the FLS method, a uniform background charge is taken into account to neutralize the system. Here, we define $x$ ($e$/cell) as the number of doped carrier per cell, where the hole/electron doping is denoted by positive/negative $e$. To see the phase transition, we calculated the total energy difference between the FM state $E(\phi = 0)$ and the ground state $E(\phi = \phi_{lowest})$ for a given $x$, as shown in Figs. 8(a) and 8(b).

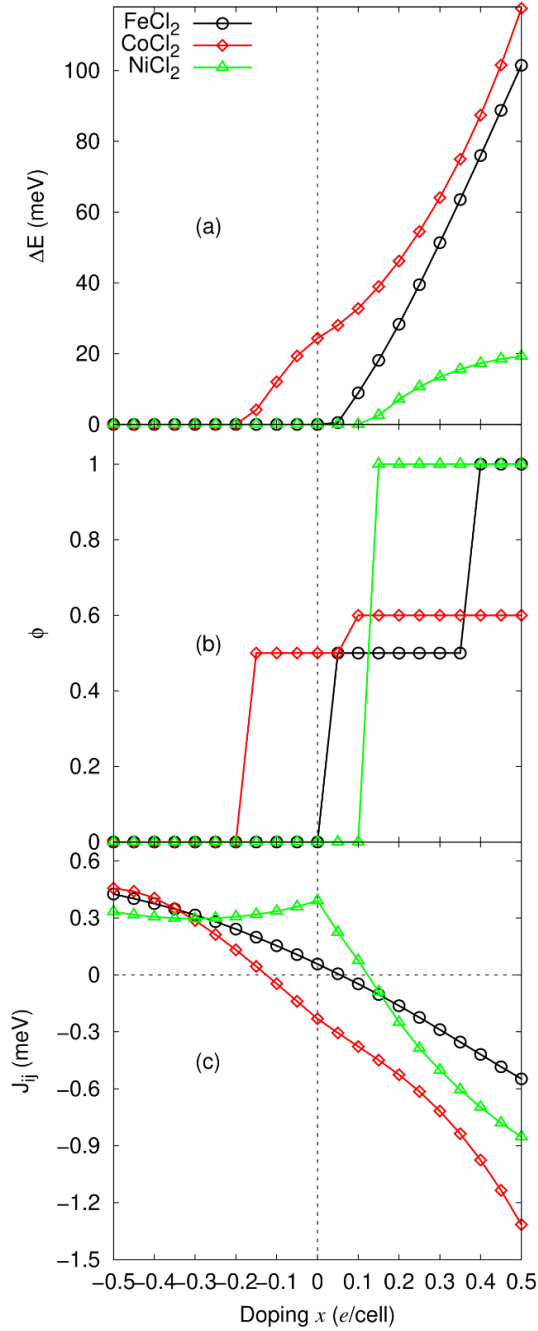

**Figure 8**. (Color online) Doping dependence of total energy difference $\Delta E = E(\phi = 0) - E(\phi = \phi_{lowest})$ (a), $\phi = \phi_{lowest}$ as a parameter in spiral vector **q** (b), and exchange parameter J$_{ij}$ (c). Here, we use the experimental lattice constants with the vacuum distance of 17.47 Å.

When we consider the hole−electron doping $x$ ($e$/cell), the other ground states emerge, thus creating a phase transition in the doping interval as shown in Figs. 8(a) and 8(b). As for the FeCl$_2$, the FM ground state still maintains in the interval of $x \leq 0$ $e$/cell, while the SP state appears in the interval of $x > 0$ $e$/cell. As for the CoCl$_2$, the SP state still holds in the interval of $x \geq -0.15$ $e$/cell, while the FM state takes part the interval of $x < -0.15$ $e$/cell. At the same time, as for the NiCl$_2$, the FM state still do exist in the interval of $x \leq 0.1$ $e$/cell, while the AFM state appears in the interval of $x > 0.1$ $e$/cell and no SP state emerges in this system. This means that spins in the magnetic atoms X for the FeCl$_2$ and CoCl$_2$ are increasingly frustrated when the hole doping increases. On the contrary, the FM state takes the domination when the electron doping increases. Therefore, the enhancement of electron in the $3d$ orbital reduces the frustrated spin to become ferromagnetic.

We also provide the dependence of exchange parameter J$_{ij}$ on the doping, as shown in Fig. 8(c). Here, we formulate $J_{ij} = 1/12(\Delta E/M^2)$ where $\Delta E = E_{\text{AFM}} - E_{\text{FM}}$ and M is the magnetic moment of the AFM state [41]. We add 1/12 to avoid the double counting on the calculation because a magnetic X atom in the 1T monolayer XCl$_2$ is surrounded by six X atoms. As we can see, the AFM direct exchange (negative J$_{ij}$) wins against the FM superexchange when the hole doping increases. For this condition, each X atom lacks the electron in the $3d$ orbital, thus allowing electron to move to the other site, thus reducing the kinetic energy due to the delocalization of electron. Nevertheless, the electron cannot hop if the spins in the neighboring X atoms are parallel. So, this spin-spin interaction causes the AFM direct exchange. In this situation, the ground state is either the AFM state or the SP state. When the electron doping increases, the FM superexchange (positive J$_{ij}$) takes the domination of magnetism. In this case, each X atom receives the electron in the $3d$ orbital, thus the Coulomb repulsion will control the occupation of electron. By the same analogy, the electron cannot hop if the spins in the neighboring X atoms are antiparallel. So, this spin-spin interaction causes the FM superexchange. In the condition the ground state can be either the FM state or the SP state.

Based on the results, we justify that the spin-spin competition creates a diversity of ground state in 2D metal dichlorides. This diversity may lead to some physical properties in these materials, such as half metallicity or topological insulating, which can be applied for future spintronic devices. In addition, introducing the doping can also change the ground state, as pointed out earlier. So, the domination of interaction, which yields a typical ground state, can be controlled by introducing the doping. At the same time, we also see that changing the lattice constant can produce different ground state.

## 4. Conclusions

We show that the ground states of monolayer 1T-XCl$_2$ (X: Fe, Co, and Ni) are sensitively determined by the lattice constant. In addition, the selected lattice constant can give significant total energy difference between the FM and AFM states, thus influencing the exchange parameter J$_{ij}$. Our findings also show that the SP state could be exist as the most stable states without any physical or chemical treatments, thus the monolayer metal dichlorides have many stable magnetic configurations. As for the SP state, the configuration can be utilized to construct the unique magnetic formations such as domain walls and skyrmion.

We also show that the phase transition, i.e., the change of ground state, can be induced by the hole−electron doping. This transition is due to the spin-spin competition between the AFM direct exchange and the FM superexchange. The competition can frustrate the spins of the magnetic atoms when they decide to choose one of those two interactions. This is due to electron hopping as the electron delocalizes over X−Cl−X that reduces the kinetic energy. This means that the magnetic ground state can be controlled by means of doping.

**Acknowledgements**
This work was performed independently without any funding sources. The detailed calculations were computed using the personal high computer at the Universitas Negeri Jakarta.
**References**

[1] T. Kurumaji, S. Seki, S. Ishiwata, H. Murakawa, Y. Tokunaga, Y. Kaneko, Y. Tokura, Magnetic-field induced competition of two multiferroic orders in a triangular-lattice helimagnet $MnI_2$, Phys. Rev. Lett. **106** (2011), 167206, https://doi.org/10.1103/PhysRevLett.106.167206.

[2] Y. Tokunaga, D. Okuyama, T. Kurumaji, T. Arima, H. Nakao, Y. Murakami, Y. Taguchi, Y. Tokura, Multiferroicity in $NiBr_2$ with long-wavelength cycloidal spin structure on a triangular lattice, Phys. Rev. B **84** (2011) 060406(R), https://doi.org/10.1103/PhysRevB.84.060406.

[3] X. Wu, Y. Cai, Q. Xie, H. Weng, H. Fan, J. Hu, Magnetic ordering and multiferroicity in $MnI_2$, Phys. Rev. B **86** (2012) 134413, https://doi.org/10.1103/PhysRevB.86.134413.

[4] J.F. Scott, Multiferroic memories, Nature Mater. **6** (2017) 256-257, https://doi.org/10.1038/nmat1868.

[5] N.A. Spaldin, M. Fiebig, The Renaissance of Magnetoelectric Multiferroics, Science **309** (2005) 391-392, https://doi.org/10.1126/science.1113357.

[6] M.A. McGuire, Crystal and magnetic structures in layered, transition metal dihalides and trihalides, Crystal **7** (2017) 121, https://doi.org/10.3390/cryst7050121.

[7] S.G. Wang, W.H.E. Schwarz, Density functional study of first row transition metal dihalides, J. Chem. Phys. **109** (1998) 7252-7262, https://doi.org/10.1063/1.477359.

[8] B.K. Rai, A.D. Christianson, D. Mandrus, A.F. May, Influence of cobalt substitution on the magnetism of $NiBr_2$, Phys. Rev. Mater. **3** (2019) 034005, https://doi.org/10.1103/PhysRevMaterials.3.03400.

[9] K.S. Novoselov, A.K. Geim, S.V. Morozov, D. Jiang, Y. Zhang, S.V. Dubonos, I.V. Grigorieva, A.A. Firsov, Electric field effect in atomically thin carbon films, Science **306** (2004) 666-669, https://doi.org/10.1126/science.1102896.

[10] K.S. Novoselov, A.K. Geim, S.V. Morozov, D. Jiang, M.I. Katsnelson, I.V. Grigorieva, S.V. Dubonos, A.A. Firsov, Two-dimensional gas of massless Dirac fermions in graphene, Nature **438** (2005) 197-200, https://doi.org/10.1038/nature04233.

[11] A.K. Geim, K.S. Novoselov, The rise of graphene, Nature Mater. **6** (2007) 183-191, https://doi.org/10.1038/nmat1849.

[12] M.-C. Wang, C.-C. Huang, C.-H. Cheung, C,-Y. Chen, S.G. Tan, T.-W. Huang, Y. Zhao,Y. Zhao, G. Wu, Y.-P. Feng, H.-C. Wu, C.-R. Chang, Prospects and opportunities of 2D van der Waals magnetic systems, Ann. Phys. **532** (2020) 1900452, https://doi.org/10.1002/andp.201900452.

[13] P. Chen, J.-Y. Zou, B.-G. Liu, Intrinsic ferromagnetism and quantum anomalous Hall effect in $CoBr_2$ monolayer, Phys. Chem. Chem. Phys. **19** (2017) 13432-13437, https://doi.org/10.1039/C7CP02158E.

[14] E.A. Kovalevaa, I. Melchakova, N.S. Mikhalevaa, F.N. Tomilina, S.G. Ovchinnikova, W. Baek, V.A. Pomogaev, P. Avramov, A.A. Kuzubov, The role of strong electron correlations in determination of band structure and charge distribution of transition metal dihalide monolayers, J. Phys. Chem. Solids **134** (2019) 324-332, https://doi.org/10.1016/j.jpcs.2019.05.036.

[15] V.V. Kulish, W. Huang, Single-layer metal halides $MX_2$ (X = Cl, Br, I): stability and tunable magnetism from first principles and Monte Carlo simulations, J. Mater. Chem. C **5** (2017) 8734-8741, https://doi.org/10.1039/C7TC02664A.

[16] A.S. Botana, M.R. Norman, Electronic structure and magnetism of transition metal dihalides: bulk to monolayer, Phys. Rev. Mater. **3** (2019) 044001, https://doi.org/10.1103/PhysRevMaterials.3.044001.

[17] T.B. Prayitno, F. Ishii, First-principles study of spiral spin density waves in monolayer $MnCl_2$ using generalized Bloch theorem, J. Phys. Soc. Jpn. **88** (2019) 104705, https://doi.org/10.7566/JPSJ.88.104705.